# Memories in the Photoluminescence Intermittency of Single Cesium Lead Bromide Nanocrystals

Lei HOU*[†,‡,¶], Chen ZHAO[¶], Xi YUAN[#], Jialong ZHAO[#], Franziska KRIEG[§,⊥], Philippe TAMARAT[†,‡], Maksym V. KOVALENKO[§], Chunlei GUO[¶], Brahim LOUNIS*[†,‡]

[†] Université de Bordeaux, LP2N, Talence, France

[‡] Institut d'Optique and CNRS, LP2N, Talence, France

[#] Jilin Normal University, Changchun, China

[§] Institute of Inorganic Chemistry, Department of Chemistry and Applied Biosciences, ETH Zürich, Zürich, Switzerland

[⊥] Empa- Swiss Federal Laboratories for Materials Science and Technology, Dübendorf, Switzerland

[¶] Changchun Institute of Optics, Fine Mechanics and Physics, Chinese Academy of Sciences, Changchun, China

*h_lei@pku.edu.cn
*brahim.lounis@u-bordeaux.fr

**Abstract:**

*Single cesium lead bromide ($CsPbBr_3$) nanocrystals show strong photoluminescence blinking, with on- and off- dwelling times following power-law distributions. We investigate the memory effect in the photoluminescence blinking of single $CsPbBr_3$ nanocrystals and find positive correlations for successive on-times and successive off-times. This memory effect is not sensitive to the nature of the surface capping ligand and the embedding polymer. These observations suggest that photoluminescence intermittency and its memory are mainly controlled by intrinsic traps in the nanocrystals. These findings will help optimizing light-emitting devices based on inorganic perovskite nanocrystals.*

**Introduction:**

Perovskite nanocrystals (NCs) of cesium lead halides (CsPbX$_3$, X = Cl, Br or I) have several attractive properties such as large absorption cross section [1], wide tunable emission range by adjusting the halide composition [2, 3], and high photoluminescence (PL) quantum yield without elaborate epitaxial overcoating for electronic passivation [4]. They also show a lack of deep traps and a high tolerance to defects [5,6]. These properties make them ideal nanomaterials for applications in LCD and LED displays [7-9], solar light energy harvesting [10,11], lasing [12,13], single-photon sources [14] and photodetection [15]. However, like most luminescent nano-emitters studied at the individual level, such as single organic molecules, fluorescent proteins and colloidal semiconductor quantum dots [16-20], they exhibit PL intermittency also called blinking, i.e. the random switching of PL intensity between high and low levels under prolonged illumination. Those dim and dark periods significantly reduce the overall PL intensity and the quantum yield, thus limiting the applications of these luminescent materials in light-emitting devices.

PL intermittency has been observed in cesium lead halide perovskite NCs [4, 21-25] over the past few years. Different models that were initially proposed for II-VI semiconductor NCs [19,26] have been invoked to rationalize the PL blinking of these NCs. For example, random charging and discharging followed by Auger non-radiative recombination has been used to explain the PL intermittency of cesium lead halide NCs at room temperature [21,27]. Diffusion-controlled electron transfer (DCET) has been proposed for the dominant charge trapping mechanism in order to explain the dependence of power-law kinetics of bright states on the excitation power [23]. Models of multiple recombination centers provide satisfactory explanations for the linear relationship between PL lifetime and PL intensity [22]. Most of the previous studies on the PL intermittency are based on the statistics of on-time and off-time durations, that is the probability density of dwelling times on the bright and dim states, and its dependence on the excitation [28, 29], the composition [21] and the temperature [30]. However, the analysis on subsequent dwelling times is largely overlooked for perovskite NCs, which was used to reveal the correlation dynamics in conventional NCs [31, 32] and single molecules [33]. In this letter, we investigate the room-temperature PL blinking of single CsPbBr$_3$ NCs by analyzing the correlation between successive dwelling times. We address the memory effect in the PL intermittency, and study its dependence on the excitation power, the nature of surface capping ligands and the embedding polymer.

**Results and discussions:**

We first examine the statistics of dwelling times on the bright and dim states in the PL blinking of single CsPbBr$_3$ NCs (see Fig. 1a. More details on their synthesis and characterization are shown in Methods and Figure S1, Supporting Information). As exemplified in Fig. 1b and e, all studied NCs exhibit pronounced PL intermittency under prolonged laser illumination. Despite random switching between bright and dim states, single CsPbBr$_3$ NCs can be fluorescent for extended times before bleach, typically several tens of minutes. The PL intensity histogram in Fig.1c shows a very broad distribution. The PL intensity autocorrelation function $g^2(\Delta t)$ of individual CsPbBr$_3$ NCs is smaller than 0.5 at zero-time delay (see Fig. 1d), which is a clear signature of photon antibunching and single photon emission. Nevertheless, the finite value of $g^2(0)$ indicates a residual contribution of biexciton emission in the PL. Under weak excitation, the ratio of quantum yields between biexciton and exciton emission can be directly estimated from the ratio of areas below the central peak and one of the lateral ones.

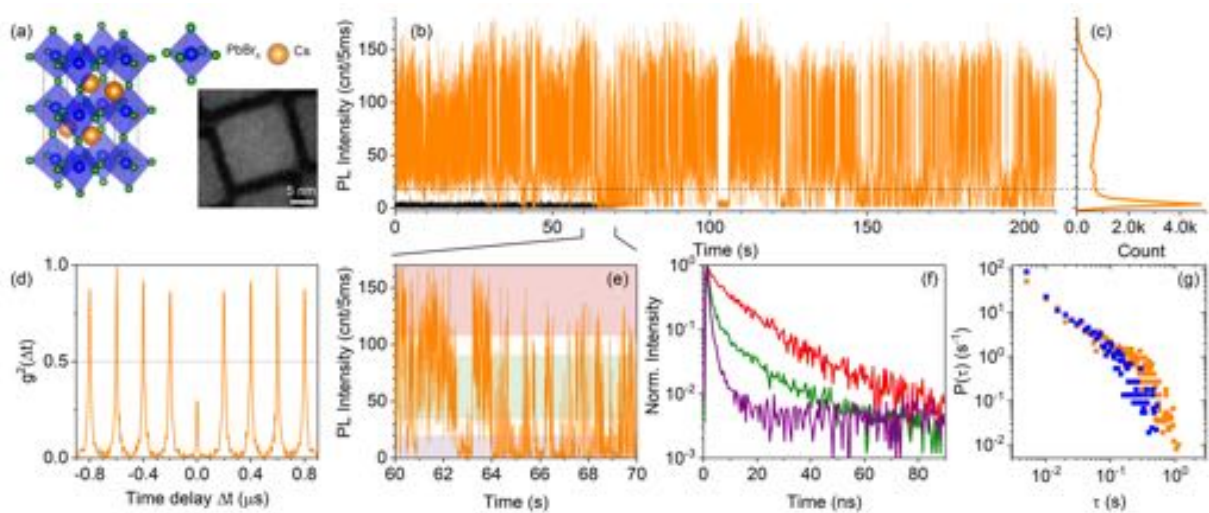

**Figure 1: Photoluminescence intermittency of single CsPbBr$_3$ NCs under prolonged illumination. (a) Cubic crystal structure of CsPbBr$_3$ NCs at room temperature. The lower insert is the high-resolution transmission electron microscopy image. (b) PL Blinking and (e) zoom-in trace under pulsed excitation at 488 nm. The background level under the same condition is shown**

in black in (b). The dashed lines show the threshold for the discrimination between the ON and OFF states. The binning time is 5 ms. (c) Histogram of PL intensities along the PL time trace. (d) PL intensity autocorrelation function $g^2(\Deltaت)$ normalized by its maximum. (f) PL decay curves at different intensity levels. Different colors correspond to the intensity levels highlighted in Fig. 1e. (g) Probability densities of on- and off- times in the traces (b). The average number of excitons under pulsed excitation is estimated to be ~ 0.06.

Taking advantage of the time-tagged time-resolved single photon data, we plot the PL decay curves at different intensity levels in Fig. 1f. Clearly, different PL intensities correspond to different decay rates, with longer (shorter) lifetime being associated with higher (lower) luminescent intensities. Dim states with lower quantum yields are often attributed to charging of the NC, which leads to the formation of a trion after the absorption of a photon [34]. The subsequent Auger recombination of the trion with a characteristic time much faster than the exciton radiative lifetime [35, 36] quenches the PL. Another possible explanation invokes the activation of reaction centers, which modifies the nonradiative decay rate of the exciton recombination [37] and alters the PL quantum yield. Indeed, the large variations of PL intensities and PL decay times along the PL time trace suggest a broad distribution of emissive states over time.

The intensity histograms are then used to define an appropriate intensity threshold to distinguish the bright (ON) and dark (OFF) periods. By setting this threshold level at the minimum between the two peaks in the intensity histogram (the dashed line in Fig. 1b), the time trace can be sectioned into a series of ON and OFF periods, whose duration will be denoted as $\tau_i^{ON}$ and $\tau_i^{OFF}$, $i$ being the index number in the series. Since the long duration events are rare compared to short ones, we weight each duration by the average time between nearest neighbor events [38]. Figure 1g shows that probability densities of on- times $P(\tau^{ON})$ and off- times $P(\tau^{OFF})$ follow a power-law distribution $P(\tau) \sim \tau^\alpha$ over a large range of times and probability densities. Similar trends are found for 30 NCs and shown in Figure S2, Supporting Information. It is noteworthy that we do not observe a significant difference in the on-time and off-time distributions between cw and pulsed excitations under moderate excitation powers (see Figure S3, Supporting Information). Therefore, we will not make a distinction between both excitation regimes in the following discussions. When the excitation power is increased, the truncation time in the on-time distribution, where the probability density deviates from the power law distribution and turns into an exponential decay at longer times, slightly shifts to smaller values (See Figure S4 , Supporting Information). Since the truncation time is often associated with the saturation of the distributed rate process [26], we deduce that charge trapping is weakly induced by light in these NCs. Based on the statistics of dwelling times discussed above, we conclude that the PL intermittency of defect tolerant $CsPbBr_3$ NCs resemble what has been observed with conventional NCs.

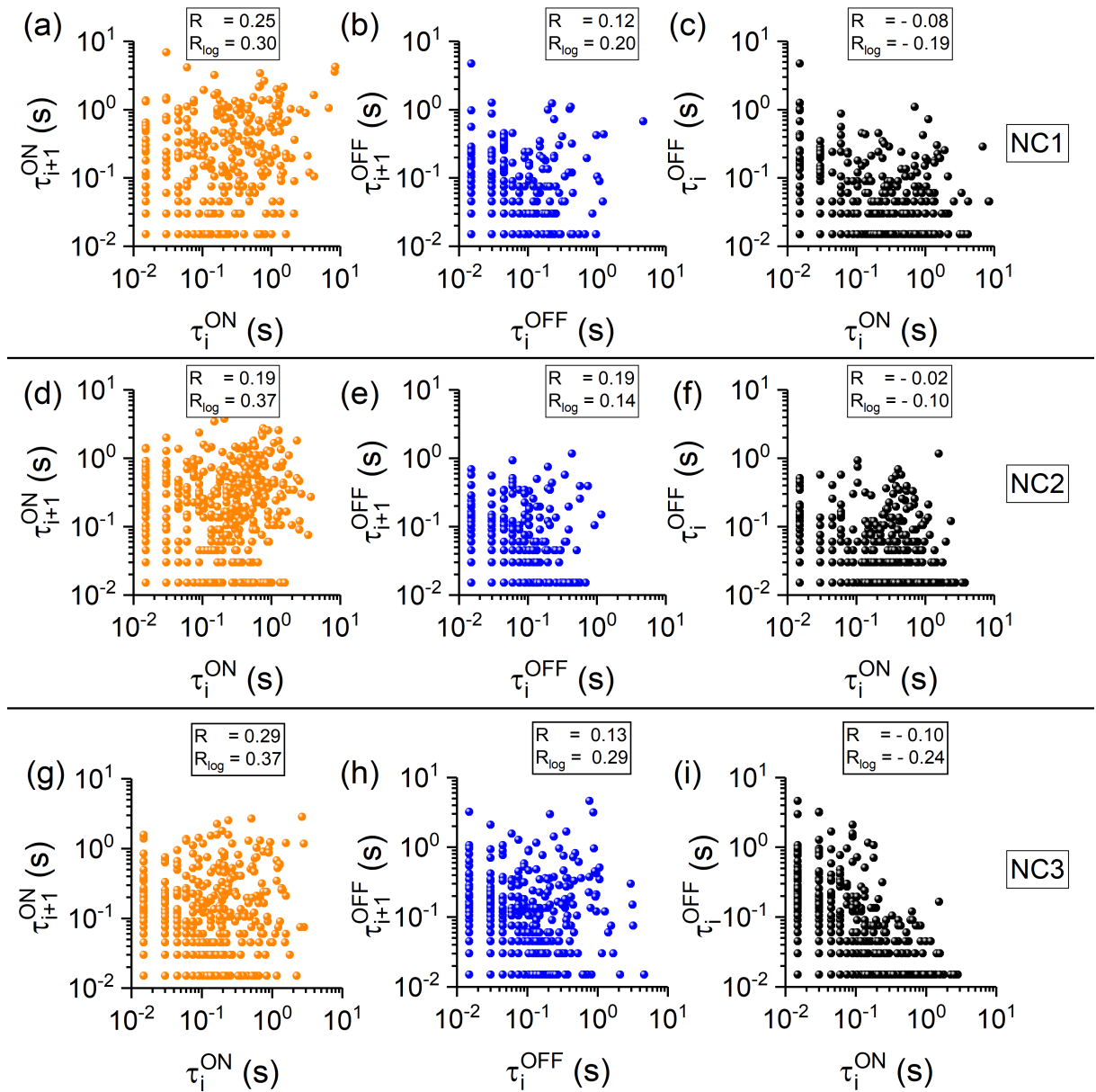

**Figure 2:** Scatter plots of successive durations extracted from experimental blinking traces. (a-c), (d-f) and (g-i) Scatter plots of successive ON times (in orange), successive OFF times (in blue) and subsequent ON-OFF times (in black) respectively, extracted from the blinking traces of three NCs NC1 (a-c), NC2 (d-f) and NC3 (g-i). The blinking time traces are shown in Figure S5, Supporting Information. The calculated correlation coefficients are listed on top of each plot. The polymer embedding the NCs is PMMA. The excitation intensity is about 9 Wcm$^{-2}$ and the binning time is 15 ms.

We now turn to the correlation between subsequent dwelling times in the PL blinking of our CsPbBr$_3$ NCs. Figure 2 shows the logscale scatter plots of $\tau_i^{ON}$ versus $\tau_{i+1}^{ON}$, $\tau_i^{OFF}$ versus $\tau_{i+1}^{OFF}$, and $\tau_i^{ON}$ versus $\tau_i^{OFF}$ respectively for three NCs. Clearly, data points of $\tau_i^{ON}$ versus $\tau_{i+1}^{ON}$ and $\tau_i^{OFF}$ versus $\tau_{i+1}^{OFF}$ in Fig. 2 tend to distribute along the main diagonal, implying a positive correlation between successive ON times and successive OFF times. The scatter plot of $\tau_i^{ON}$ versus $\tau_i^{OFF}$ seems in favor of the ON times and shows a small extent of anti-correlation. To quantify these correlations, we have calculated the Pearson correlation coefficient R, which is defined as:

$$R(m) = \frac{\sum_{i=1}^{N}(\tau_i^A - \overline{\tau^A})(\tau_{i+m}^B - \overline{\tau^B})}{\sqrt{\sum_{i=1}^{N}(\tau_i^A - \overline{\tau^A})^2}\sqrt{\sum_{i=1}^{N}(\tau_{i+m}^B - \overline{\tau^B})^2}}$$

where A, B $\in \{ON, OFF\}$, N is the total number of dwelling time durations on the ON or OFF state. We have also calculated the coefficient R$_{log}$ which expresses as R, replacing times $\tau$ by $log_{10}(\tau/\tau_0)$, where $\tau_0$ is the binning time in our experiments. The R and R$_{log}$ values will be used as a measure of the correlation strength in our following discussions. The largest (smallest) value approaching +1 (-1) indicates total correlation (anti-correlation) between dwelling times. The values obtained for successive dwelling times ($m$ =1) for the three NCs confirm the correlations observed in the scatter plots and suggest a memory effect in their luminescence blinking. Positive correlations for successive $\tau^{ON}$ and successive $\tau^{OFF}$ are observed on all studied (~76) NCs, as shown in the histograms of R and R$_{log}$ presented in Figure S6, Supporting Information. For comparison, we use a Monte Carlo method to simulate an intensity time trace of a blinking emitter where no correlation exists between subsequent dwelling times. The time durations are generated randomly and follow an inverse power-law distribution. They are then used to build the scatter plots and to calculate the correlation coefficients (see Figure S7, Supporting Information).

The simulated scatter plots clearly differ from the experimental ones and the associated correlation coefficients vanish, which further suggests that a memory effect does exist in the PL blinking of single CsPbBr$_3$ NCs.

The memory observed here can be captured by the multiple recombination centers (RCs) model [37, 39]. In this model, the PL blinking is attributed to the presence of light-activated RCs or traps. The activation or deactivation of these RCs, which occurs over widely distributed time intervals, opens or closes additional non-radiative decay channels for the relaxation of excitons. As a result, the PL intensity switches between high and low levels over time. In this scenario, the longest ON time corresponds to the case where all the RCs are inactive. The PL of the NC starts to be quenched when some of the fast switching RCs are activated. It is likely that some of the fastest RCs switch back to their inactive state and the NC becomes bright again for a long period of time. This explains the positive correlation between successive ON times, and the observation that a long ON time is accompanied by short OFF times. An analogous reasoning applies for the positive correlations between successive OFF time durations.

In order to see the evolution of this memory effect, we plot the correlation coefficients as functions of the number of switching cycle $m$ in Fig. 3. We find that the correlation coefficients $R(m)$ and $R_{log}(m)$ show a similar decay with $m$. The memory is lost on average after about 80 switching cycles, where both R and R$_{log}$ values approach zero. A characteristic minimum time of this memory effect can be extracted from $m$, assuming that all switching cycles have the same duration and taking the minimum time elapsed between On and OFF periods in the PL blinking time trace. Since the intermittency times follows a power-law distribution, the binning time sets this minimum (2 × 15ms). Therefore, one can estimate that the blinking memory lasts at least 2 s. A similar value was found for CdSe NCs [31, 32], which indicates a similarity in the memory dynamics between the two types of NCs.

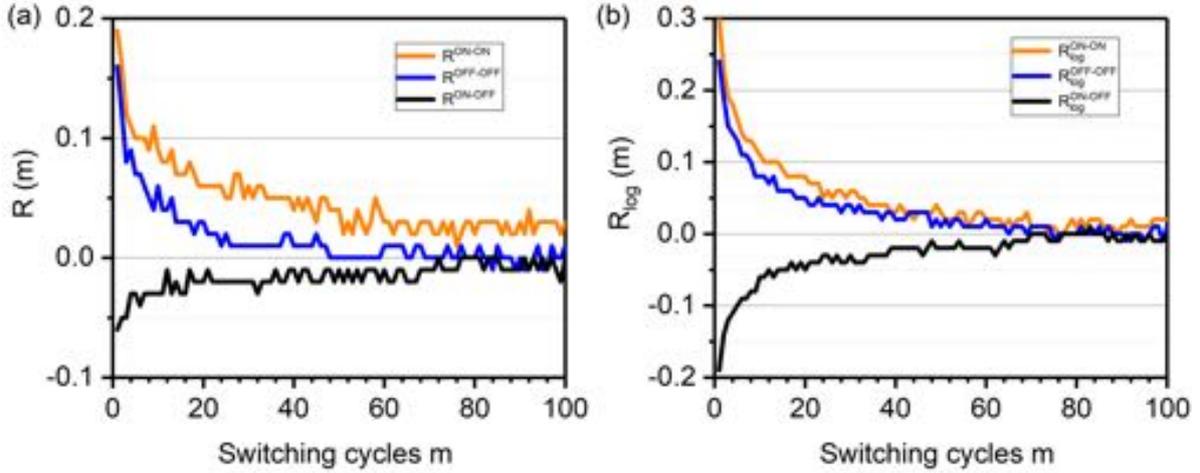

**Figure 3: Correlation coefficients decay with the number of ON-OFF switching cycles. (a) R values. (b) $R_{log}$ values. Different colors are used to distinguish the correlation for $\tau_i^{ON}$ versus $\tau_{i+m}^{ON}$ (orange), $\tau_i^{OFF}$ versus $\tau_{i+m}^{OFF}$ (blue), and $\tau_i^{ON}$ versus $\tau_{i+m}^{OFF}$ (black). The curves are obtained by averaging the correlation coefficients extracted from 76 blinking traces of single NCs.**

In the multiple RCs model, RCs or traps are supposed to be light-activated [23, 37]. We have thus varied the excitation power to seek changes in the correlation between subsequent times. We show the correlation coefficients of four single CsPbBr$_3$ NCs as a function of the cw power in Figure S8, Supporting Information. Overall, no significant power dependence of the correlation coefficients is observed, which suggests that the memory effect in the PL blinking is not sensitive to the excitation power. One can however distinguish a slight increase in the successive ON-time correlations with the power. A similar power dependence was also found with CdSe NCs [30, 31], which confirms the common blinking features shared by the perovskite and conventional NCs.

It has been shown that PL intermittency of CdSe NCs can be influenced by the conformational changes of surface capping ligands and the surrounding medium [40, 41]. In order to locate the traps responsible for PL blinking in CsPbBr$_3$ NCs and to clarify their implication in the memory effect, various NCs surface coating ligands and embedding polymers (with different dielectric permittivity) have been systematically studied. The influence of surface capping ligands on the successive time correlations is studied on CsPbBr$_3$ NCs with 10 nm size, using two types of covering ligands: oleylamine (OAm) mixed with oleic acid (OA) and zwitterionic 3-

(N,N-dimethyloctadecylammonio)-propanesulfonate molecules (see Figure S9, Supporting Information, for the molecular structures). Because of the capability of coordinating simultaneously to the surface cations and anions, zwitterionic molecules are expected to bind stronger to the NCs as compared to the conventional ammonium capping ligands [3]. Indeed, ensemble measurements have shown that zwitterionic molecules can improve the chemical durability of lead halide perovskite NC colloids: stock solution of $CsPbBr_3$ NCs passivated with these ligands can be fluorescent for more than half a year, while NCs with OAm ligands becomes non-fluorescent shortly after one month.

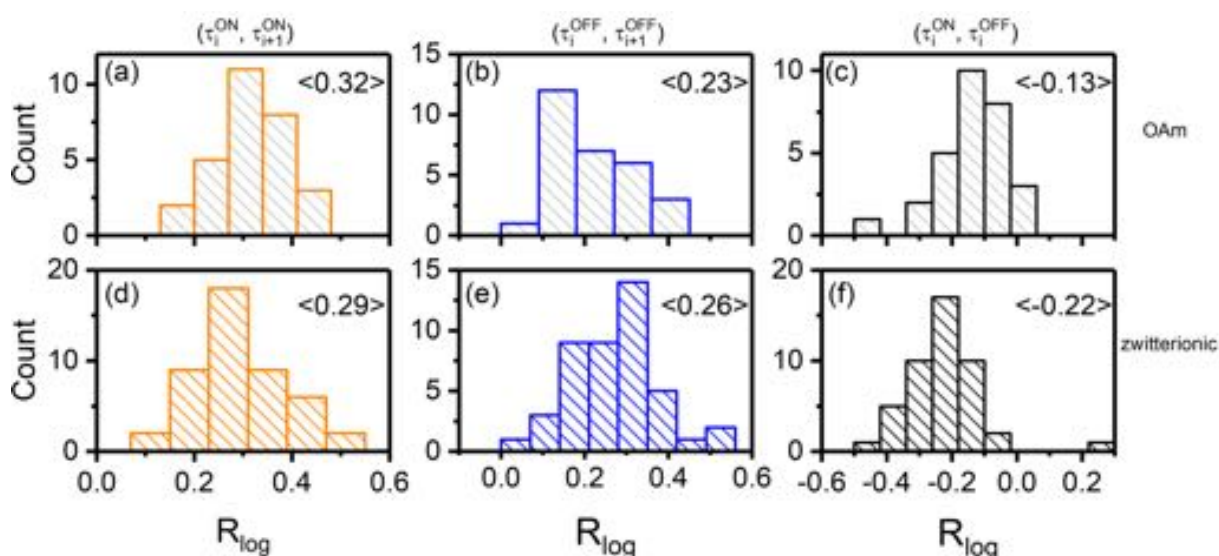

**Figure 4: Correlation coefficient $R_{log}$ between two different ligands capping the $CsPbBr_3$ NCs. (a, b, c) OAm ligands; (d, e, f) Zwitterionic ligands. Average $R_{log}$ values are indicated in brackets. Correlation pair names are listed on the top of the figures. The embedding polymer is PMMA. The NCs have an average size about 10 nm.**

In Figure 4, we show the correlation coefficient $R_{log}$ for two different ligands capping the NCs. Histograms of R values can be found in Figure S10, Supporting Information. An influence of the capping ligands to the PL blinking and its memory effect can be expected if these processes were related to the conformational reorganization of surface covering molecules. However, no significant difference in the memories is found when using the two ligands. This suggests that the traps responsible for the PL blinking of single CsPbB3 NCs do not originate from the passivation ligands.

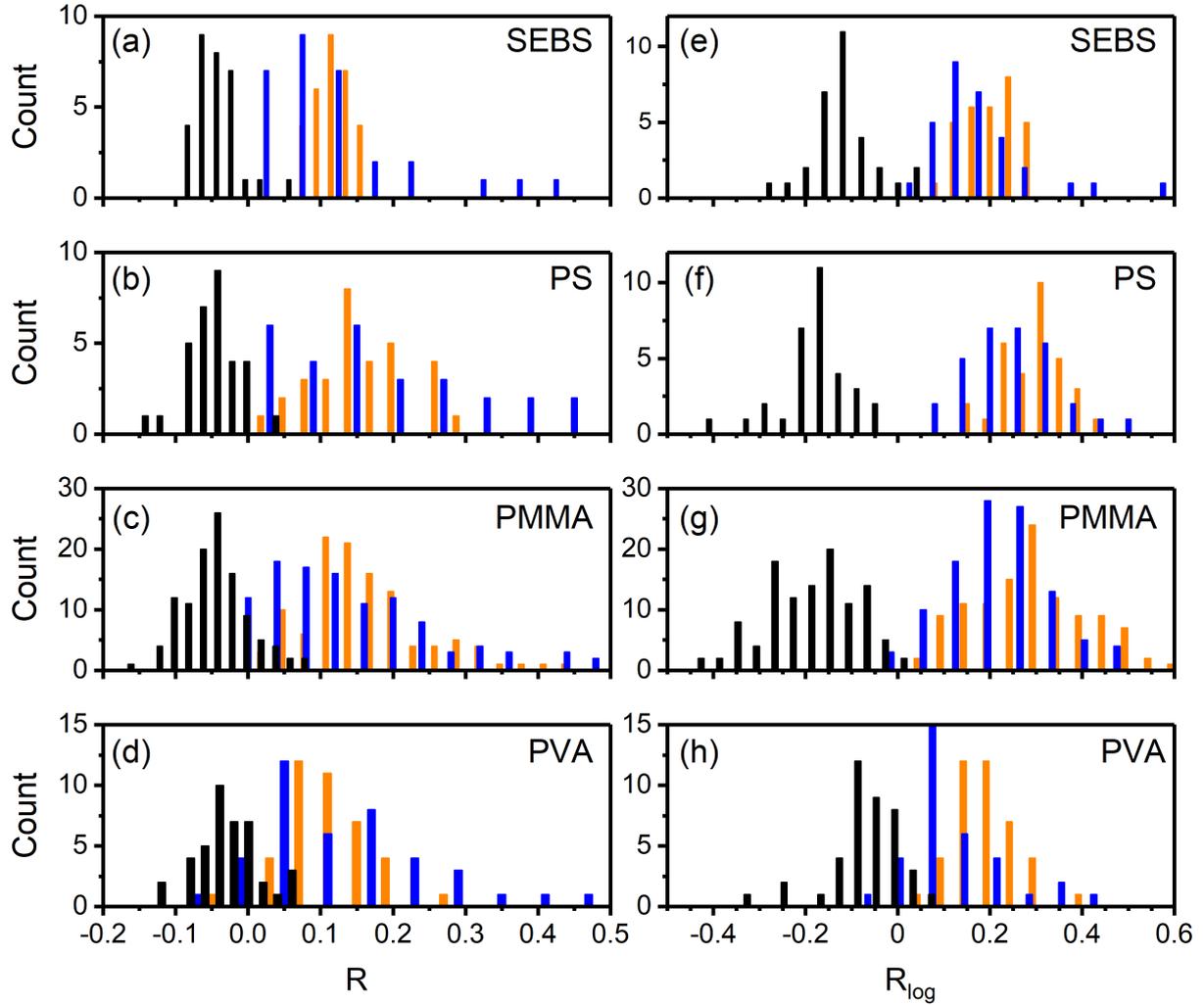

**Figure 5: Dependence of the memory on the embedding polymers. The average size of CsPbBr$_3$ NCs used in this study is about 6 nm. The experiment is done in the wide-field configuration, with an excitation intensity of about 35 Wcm$^{-2}$. The surface ligands are zwitterionic molecules. (a-d) R values between successive ON times (in orange), successive OFF times (in blue) and subsequent ON-OFF times (in black); (e-h) R$_{log}$ values. The names of the embedding polymers are listed on the figures.**

We then investigate the dependence of PL intermittency on the embedding polymer. Previous studies have shown that PL blinking of single CdSe NCs depends on the dielectric permittivity of the embedding polymer [41, 42], as the photo-induced charge rejection and its subsequent trapping by the surface can be stabilized by the surrounding medium. Four polymers matrices are chosen in this study: Styrene-ethylene-butylene-styrene (SEBS), Polystyrene (PS), Poly(methyl methacrylate) (PMMA) and Poly(vinyl alcohol) (PVA). The relative dielectric permittivity of these polymers is shown in Figure S11, Supporting Information. From the correlation coefficients shown in Fig. 5, we did not observe any clear dependence of the distributions of R and $R_{log}$ on the embedding polymer. In addition, ON and OFF dwelling time distributions in Figure S12, Supporting Information, also show the same trend in different polymers. These results lead us to the conclusion that charge rejection and its subsequent trapping in polymers does not apply to the PL blinking of single $CsPbBr_3$ NCs, and the traps are not within the embedding polymers. Additional measurements with different NC sizes in various polymer matrices give similar results (see the Fig. S13, Supporting Information), suggesting that the traps responsible for the PL blinking properties in these NCs are intrinsic to the perovskites rather than induced by the embedding environments.

**Conclusions:**

We have examined the correlation and memory effect in the PL intermittency of single $CsPbBr_3$ NCs. A positive correlation and memory are found between successive ON times and successive OFF times. Such memory is not sensitive to conditions such as the excitation power, the type of surface capping ligands, the embedding polymer and NC sizes. We deduce that the recombination centers or traps that are responsible for the PL intermittency originate from the light-induced ionic reorganization within the nanocrystal or at its surface. The candidate ions are unpaired lead ions and/or bromide vacancies, which are commonly observed as point defects in perovskite materials [43]. Indeed, experimental reports have shown that the PL quantum yield of $CsPbBr_3$ NCs can be significantly improved by bromide precursors enrichment [44] or repairing the lead-rich surface [45]. Future optimization on the light-emitting devices based on inorganic perovskite NCs should focus on eliminating these intrinsic defects or traps present in the NCs.

**Methods:**

Preparation of colloidal $CsPbBr_3$ NCs:

$CsPbBr_3$ NCs with oleylamine (OAm) and oleic acid (OA) capping ligands were synthesized according to the protocols from Ref. [46]; $CsPbBr_3$ NCs with zwitterionic capping ligands were synthesized using the method in Ref. [3]. Ensemble absorption and PL spectra, together with scanning electron microscope (SEM) images of the NCs, are shown in Figure S1, Supporting Information.

Isolated CsPbBr$_3$ NCs embedded in a polymer were prepared as follows. For PMMA and PS polymers soluble in chloroform and toluene, stock colloid solution was diluted to a concentration about 100 pM with 10 mg/mL polymer solutions. Then a droplet of the diluted solution was spin-coated on clean cover glass. For polymer PVA which is not soluble in toluene and chloroform, PVA was first dissolved in water with a concentration 10 mg/mL and spin-coated on glass. Then stock colloid solution diluted with chloroform was spin-coated on top of the PVA film. For the measurements, the sample was covered by another clean cover slip coated with a PVA film to form a sandwich structure. This difference in sample preparation can lead to a loose contact of NCs with PVA compared with other polymers. For the small size CsPbBr$_3$ NCs with average edge length of about 6 nm, cyclohexane was used as a good solvent for polymer SEBS, PS and PMMA.

Optical setup:

We used both confocal and wide-field microscopy in our experiments. The confocal set-up contains a diode laser that can deliver both continuous and picosecond pulsed light at the wavelength of 488 nm. PL emission of single CsPbBr$_3$ NCs was collected with a high NA immersion objective, separated from excitation by 488nm long pass filters. A 50:50 beam-splitter is used in the detection path to split the emission into two single-photon avalanche photodiodes in the Hanbury Brown and Twiss configuration. Time-correlated single-photon counting (TCSPC) data were recorded by a TimeHarp260p card in the time-tagged time-resolved mode. The instrument response function (IRF) is about 100 ps for the TCSPC measurements.

In order to study a large number of single NCs, we used wide-field microscopy to study the PL blinking of many individual CsPbBr$_3$ NCs. The home-built optical set-up is based on an inverted microscope. The PL emission was filtered using a bandpass filter 520/35 and was detected with an EMCCD detector (PRINCETON ProEM 512B). Typical exposure time per frame is 15 ms, and 20,000 frames are normally recorded under cw illumination. PL intensity traces as a function of illumination time were extracted with the open-source software "iSMS [47]".

**Acknowledgements:**

The authors thank Dr. Frank Krumeich (ETH Zurich) for additional TEM measurements. We acknowledge the financial support from the French National Agency for Research, Région Aquitaine, Idex Bordeaux (LAPHIA Program). B.L. acknowledges the Institut universitaire de France.

**TOC:**

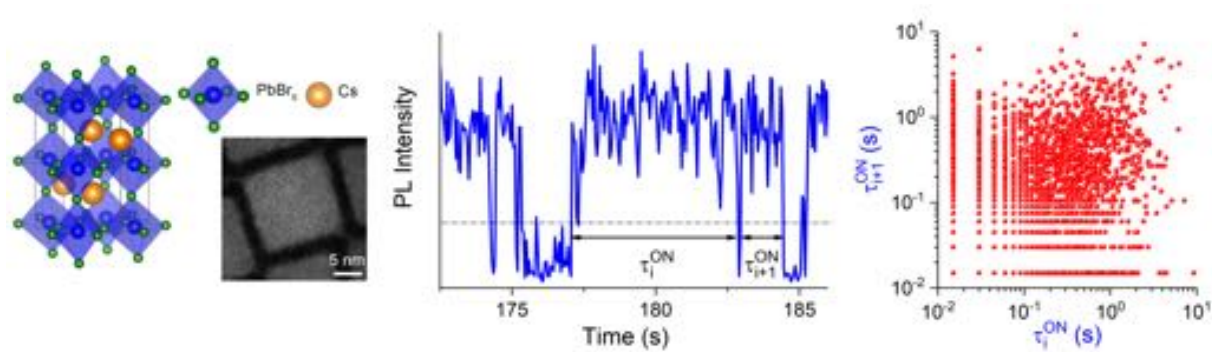

Supplementary Information for

**Memories in the Photoluminescence Intermittency of Single Cesium Lead Bromide (CsPbBr$_3$) Nanocrystals (NCs)**

S1. Absorption and PL spectra of CsPbBr$_3$ NCs colloids and TEM images;

S2. Dwelling time distributions of single CsPbBr$_3$ NCs with OAm capping ligands in PMMA polymer;

S3. Comparison of PL intermittency under cw and pulsed excitations;

S4. Excitation power dependence of ON and OFF dwelling time distributions;

S5. PL blinking traces and intensity histograms of three CsPbBr$_3$ NCs mentioned in the main text;

S6. Histograms of correlation coefficients R and R$_{log}$ for successive dwelling times;

S7. Blinking simulations with a Monte Carlo method;

S8. Excitation power dependence of the correlation coefficients R and R$_{log}$ for four CsPbBr$_3$ NCs embedded in PMMA polymer;

S9. Chemical structures of the two surface capping ligands used in our studies;

S10. Histograms of R-values for different ligands capping the NCs;

S11. Table of the relative dielectric permittivity of different polymers;

S12. Dwelling time distribution for NCs embedded in different polymers;

S13. Correlation coefficient histograms for NCs of various sizes in various polymers.

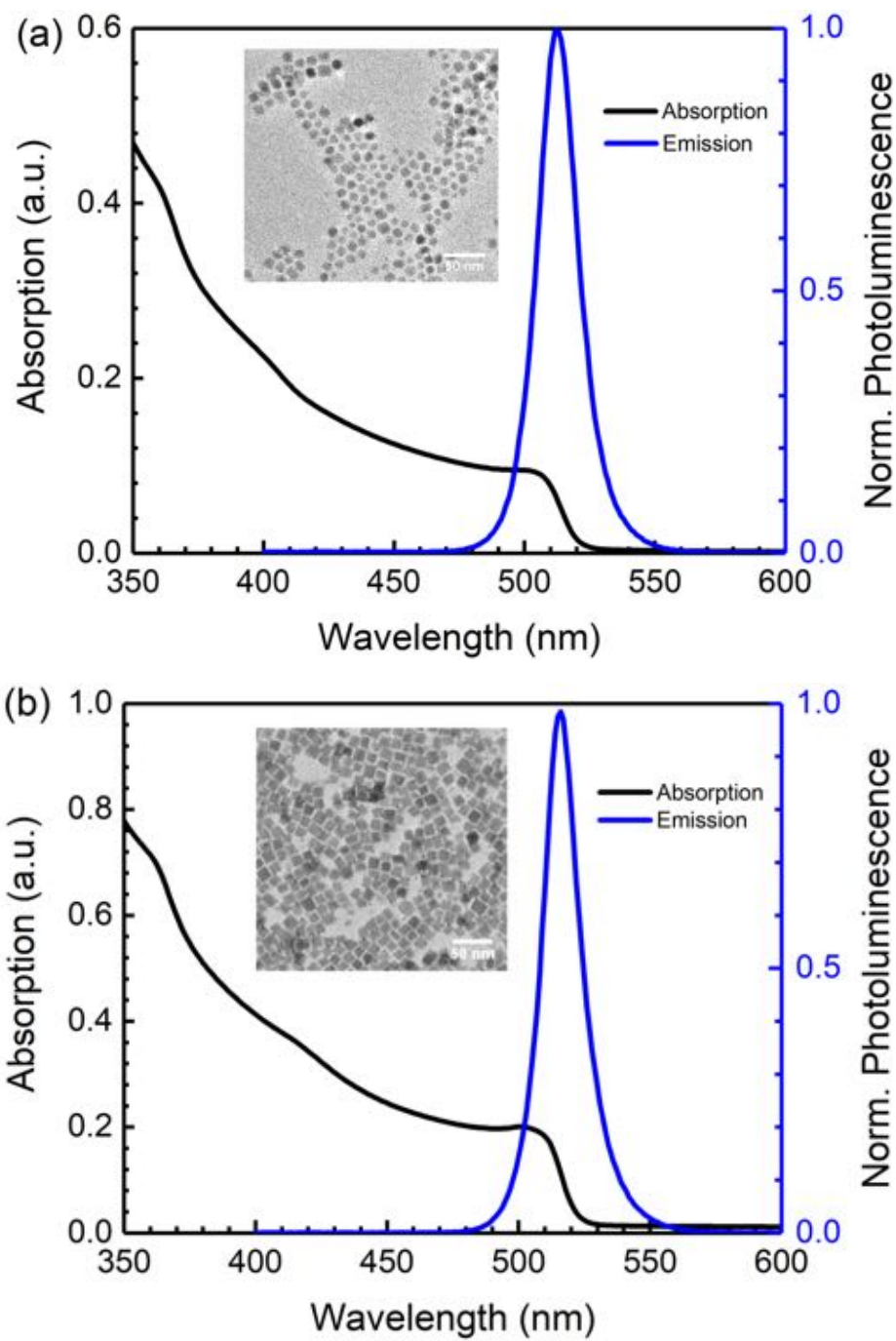

**Figure S1: Absorption and PL spectra of CsPbBr$_3$ NCs colloids and TEM images.** (a) Spectra of colloidal CsPbBr$_3$ NCs with OAm/OA capping ligands in toluene. The scale bar in the insert TEM image is 50 nm. The average size is 10.2 nm. (b) Spectra of colloidal CsPbBr$_3$ NCs with zwitterionic capping ligands in toluene. The average size measured over 76 NCs is 10 nm.

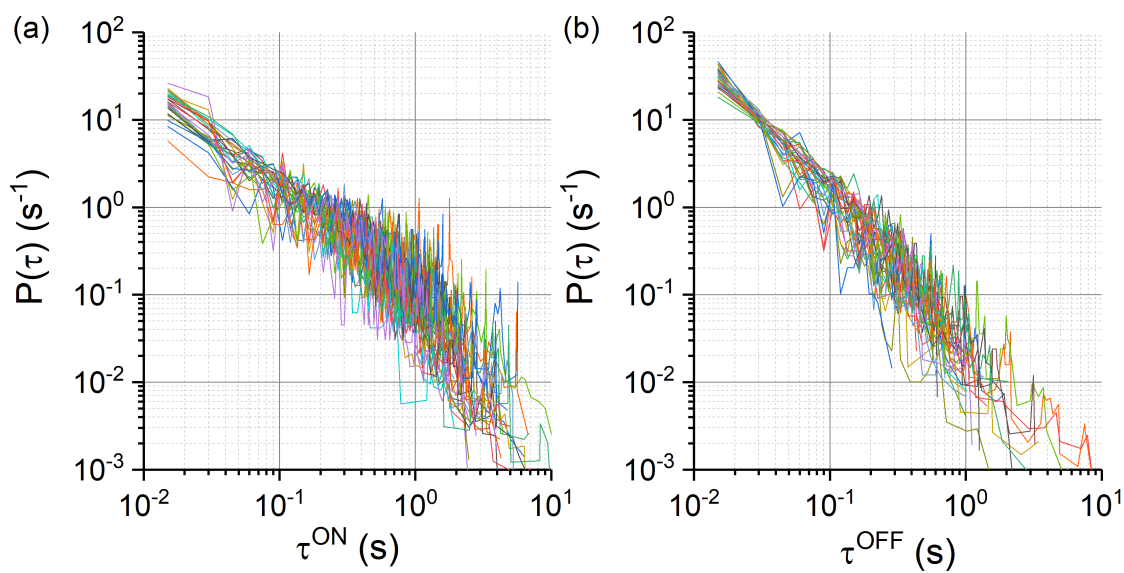

**Figure S2: Dwelling time distributions of single CsPbBr3 NCs with OAm capping ligands in PMMA polymer.** (a) ON and (b) OFF time distributions plotted with different colors. The time bin is 15 ms.

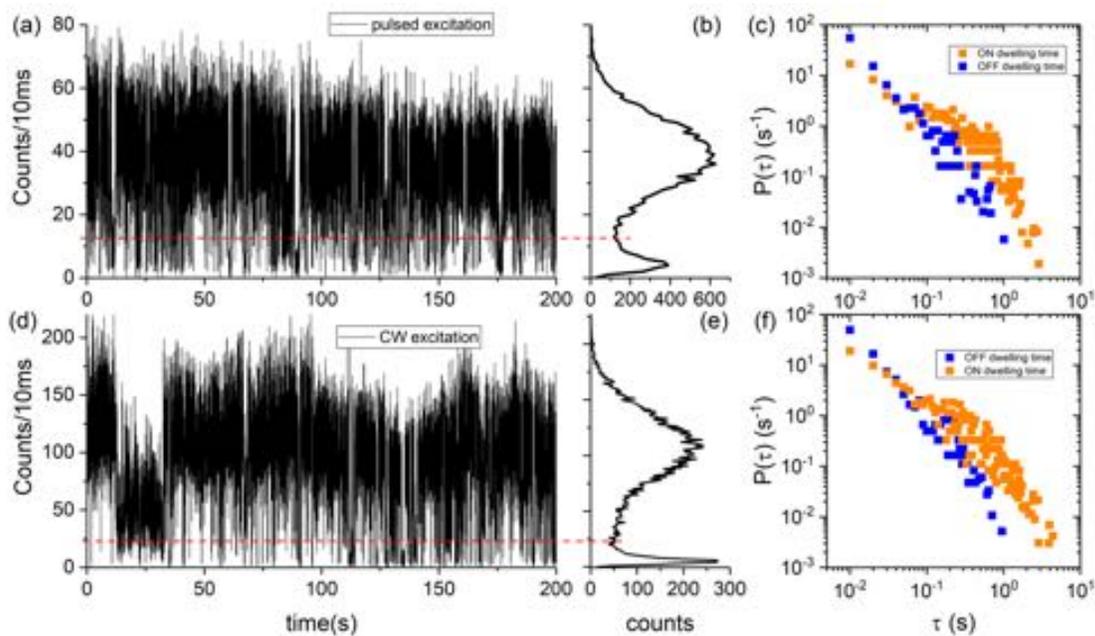

**Figure S3: Comparison of PL intermittency under cw and pulsed excitations.** (a-c) under pulsed excitation at 488 nm; (d-f) under cw excitation. The experiments were done on the same NC. The distributions of dwelling times are similar in both regimes of excitation.

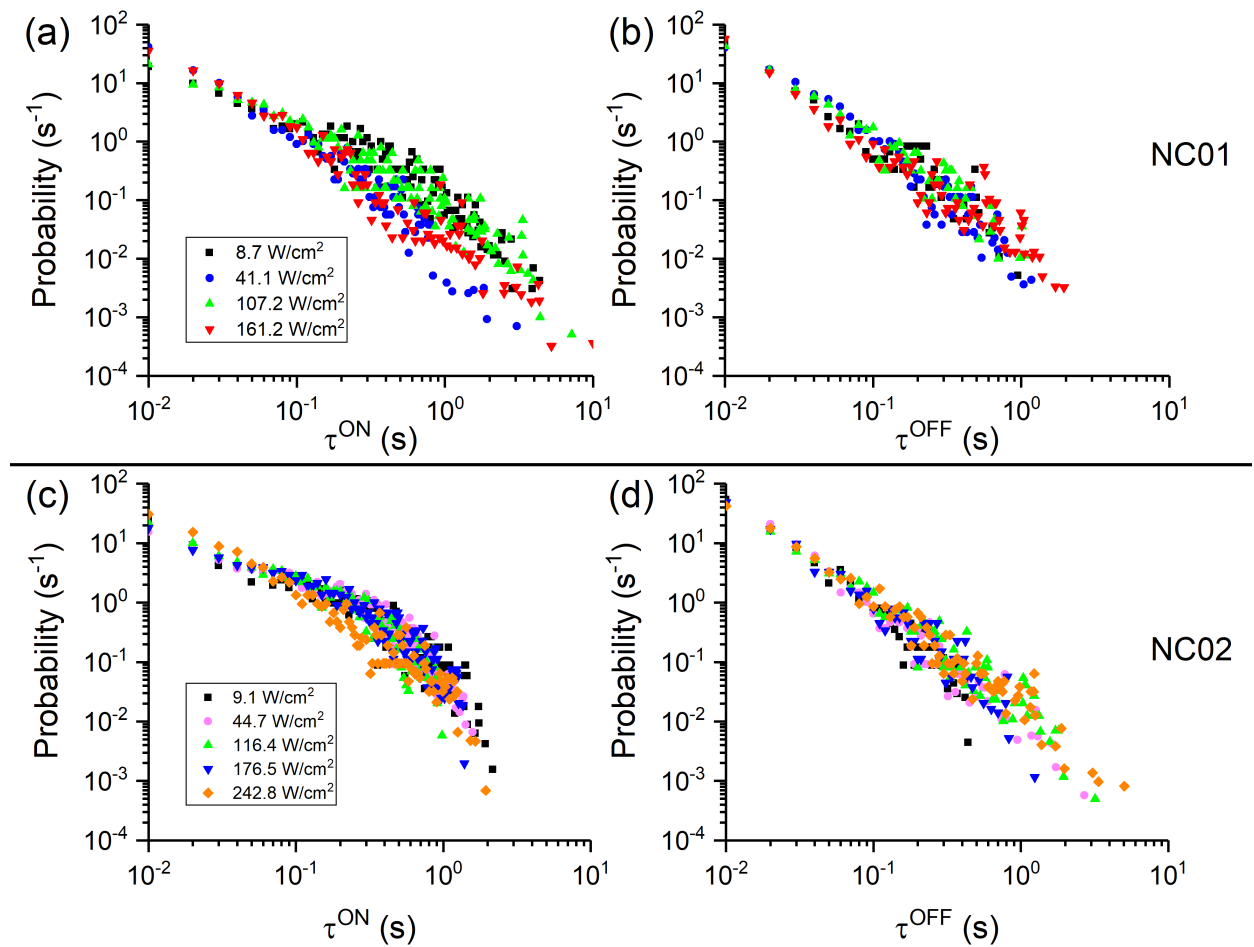

**Figure S4: Excitation power dependence of ON and OFF dwelling time distributions.** Data of two CsPbBr$_3$ NCs (NC01 and NC02) are shown. (a, c) ON time probability densities; (b, d) OFF time probability densities. The binning time is 10 ms for both NCs.

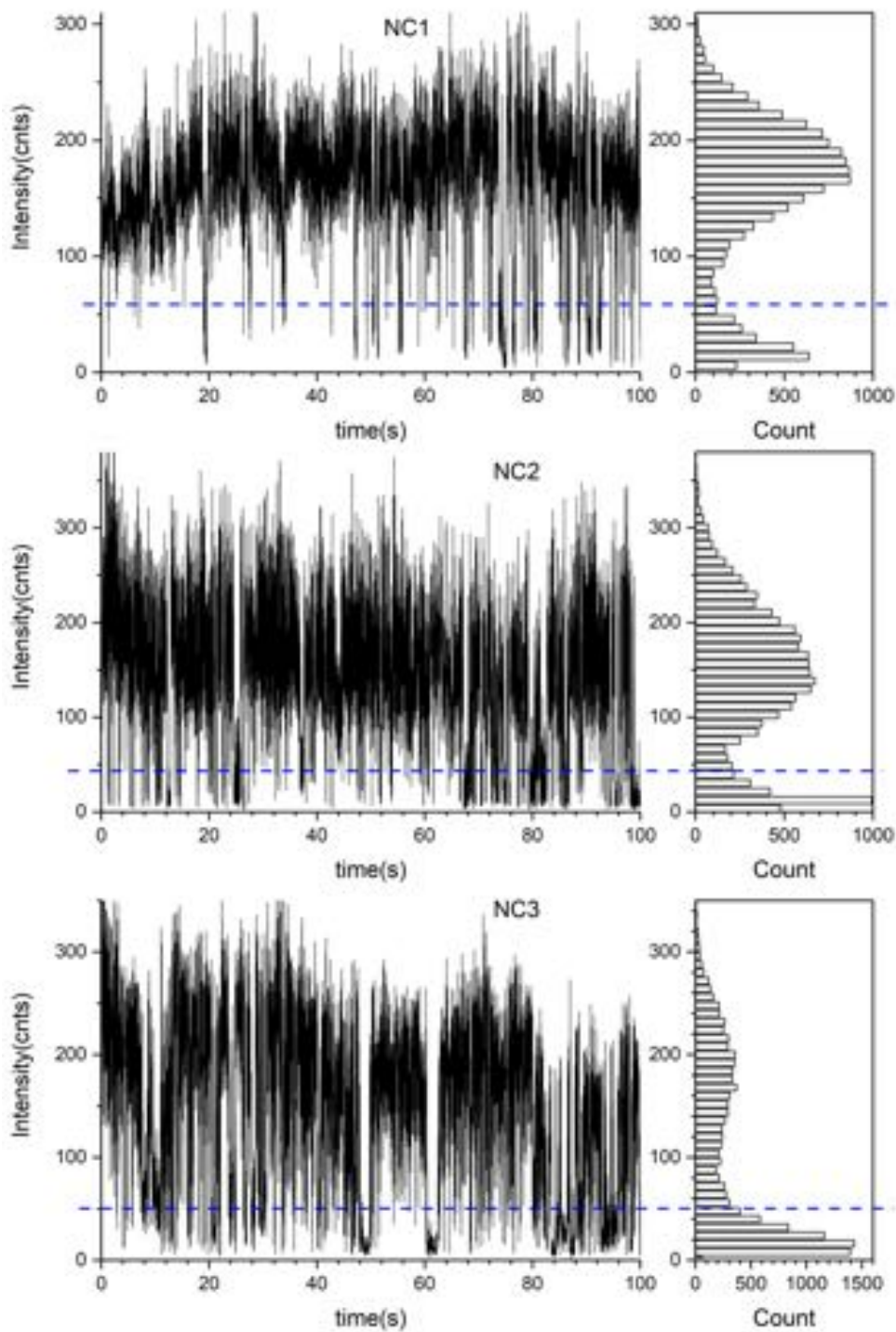

**Figure S5: PL blinking traces and intensity histograms of three CsPbBr$_3$ NCs mentioned in the main text.** The Binning time is 15 ms. The excitation power is about 8 W cm$^{-2}$ at 488 nm.

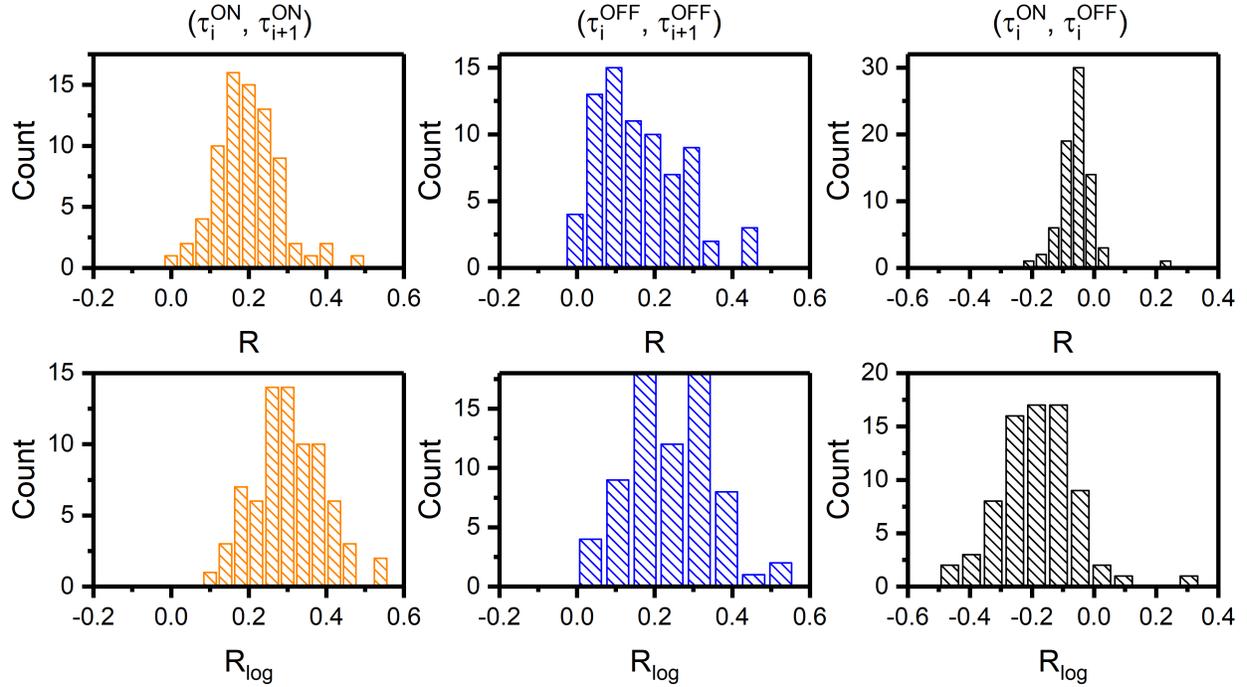

**Figure S6: Histograms of correlation coefficients R and $R_{log}$ for successive dwelling times.** Successive ON times (orange), successive OFF times (blue) and subsequent ON-OFF times (black). The histograms are built with 76 NCs' data. The distribution in R and Rlog values reflects the inhomogeneity in NCs' sizes or in the local environments.

In the histograms shown here and in the main text, there are rare cases where positive $\tau_n^{ON}$ versus $\tau_n^{OFF}$ correlation or negative $\tau_n^{OFF}$ versus $\tau_{n+1}^{OFF}$ correlation were observed, which is most likely caused by the limitation of the threshold method and the overlapping region in the PL intensity histogram. The accuracy of such correlation analysis indeed depends on the threshold level to discriminate the ON and OFF states as well as the binning time of the blinking trace (15 ms in our analysis) [1,2]. In order to minimize such influences, we choose blinking traces presenting two well-defined peaks in the intensity histogram. Stefani et.al have examined these influences and concluded that as long as the maximal Poisson-distributed OFF intensity can be identified, the influence of threshold on the correlation coefficients is negligible [3].

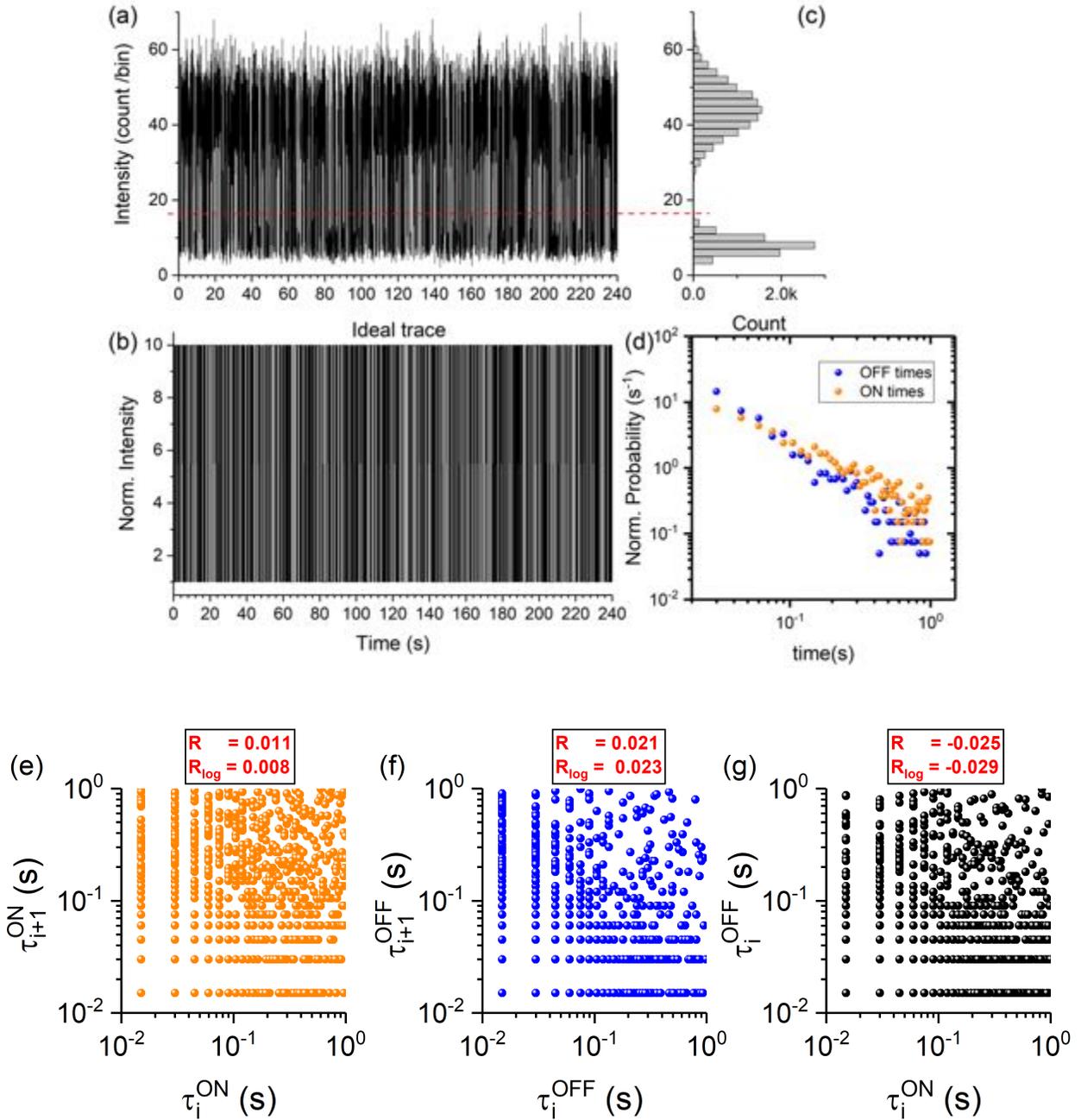

**Figure S7: Blinking simulations with a Monte Carlo method.** (a) Simulated time trace, together with the intensity histogram (c). (b) Ideal time trace composed of randomly generated time intervals. The randomly generated time intervals with value 1 and 10 follow inverse power-law statistics with power components of -1.09 (ON) and -1.47 (OFF) respectively. The reason of choosing these values is based on the average power-law components of 29 NCs. (d) The ON (in orange) and OFF (in blue) time probability densities when the threshold to distinguish ON and OFF states is chosen at the red dashed line. (e, f, g) scatter plots of successive ON times (in orange), successive OFF times (in blue) and subsequent ON-OFF times (in black). The calculated correlation coefficient R and $R_{log}$ values are shown on the top of each plot and have negligible values compared with those of NCs.

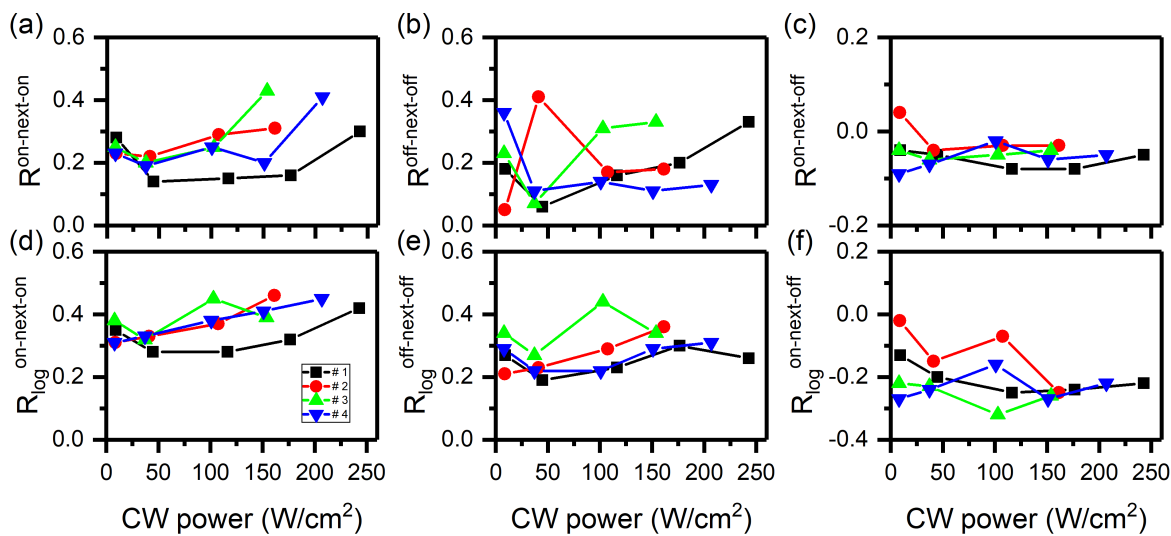

**Figure S8: Excitation power dependence of the correlation coefficients R and $R_{log}$ for four CsPbBr$_3$ NCs embedded in PMMA polymer.** Each color represents a single NC. (a, b, c) R values; (d, e, f) $R_{log}$ values.

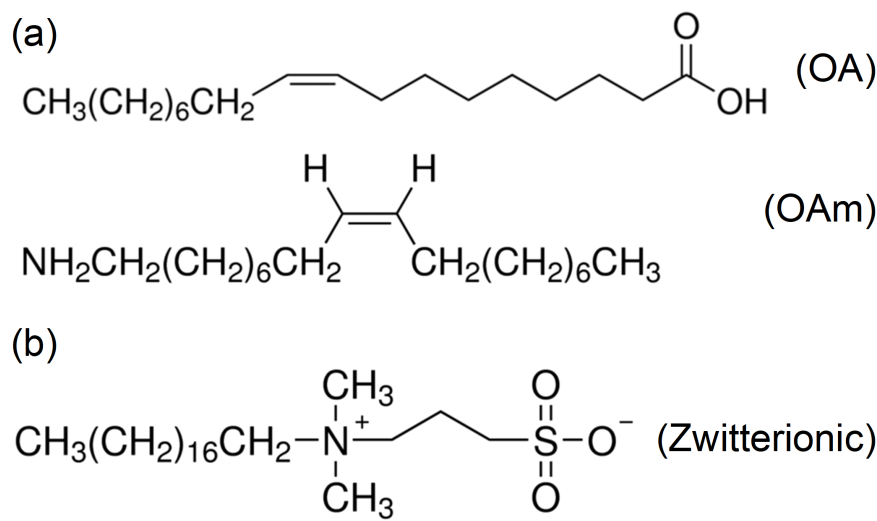

**Figure S9: Chemical structures of the two surface capping ligands used in our studies.** (a) Oleic acid (OA) and oleylamine (OAm) ligands. In the precursor for the synthesis of CsPbBr$_3$ colloid, OA is the anion and OAm is the cation. The OAm will bond to the bromide-rich surface and serve as passivation layers for the NCs. (b) Zwitterionic molecule.

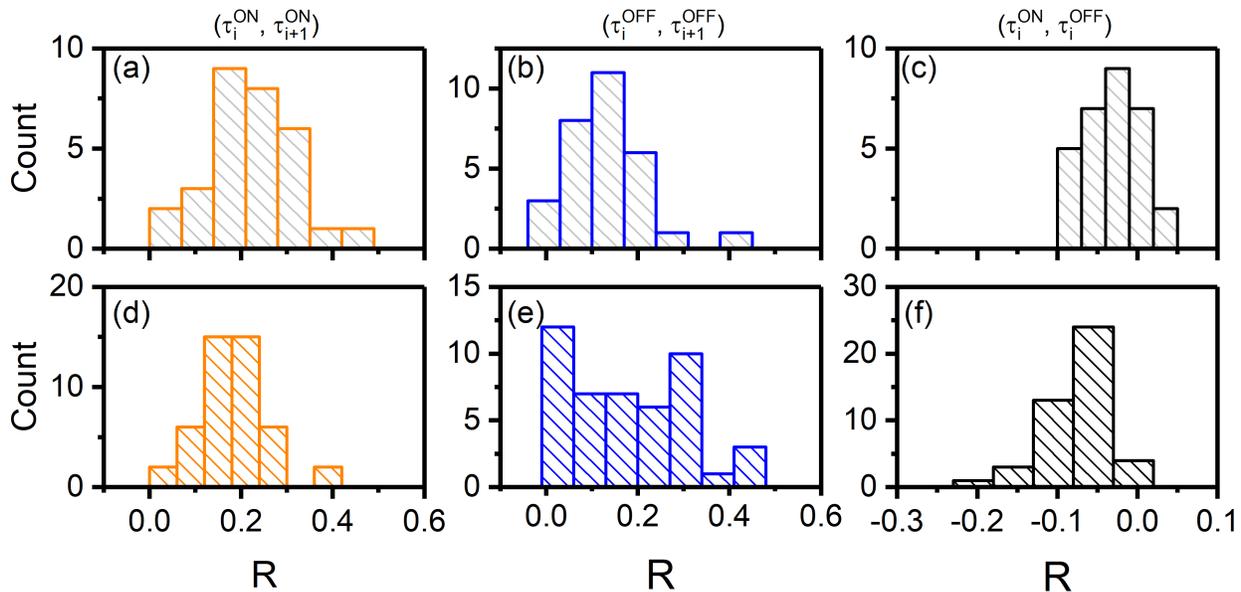

**Figure S10: Histograms of R-values of subsequent periods for different ligands capping the NCs.** (a, b, c) R values for NCs with OAm ligands; (d, e, f) R values for NCs with zwitterionic capping ligands. (a, d), (b, e), (c, f) were calculated with successive ON times, successive OFF times and subsequent ON-OFF times, respectively.

|  | SEBS | PS | PMMA | PVA |
|---|---|---|---|---|
| $\varepsilon_r$ | 2.2 | 2.53 | 3.4 | 14 |

**S11: Table of the relative dielectric permittivity of different polymers.**

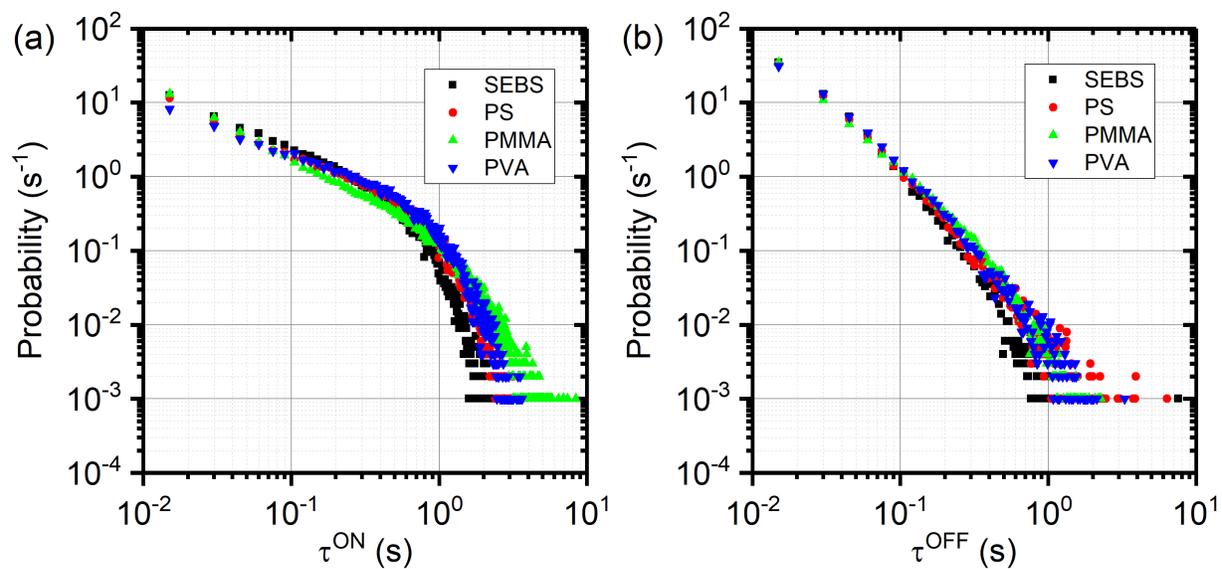

**Figure S12: Dwelling time distribution for NCs embedded in different polymers.** (a) ON time distribution; (b) OFF time distribution. The points are built from average results over a hundred single NCs with a size of ~6 nm.

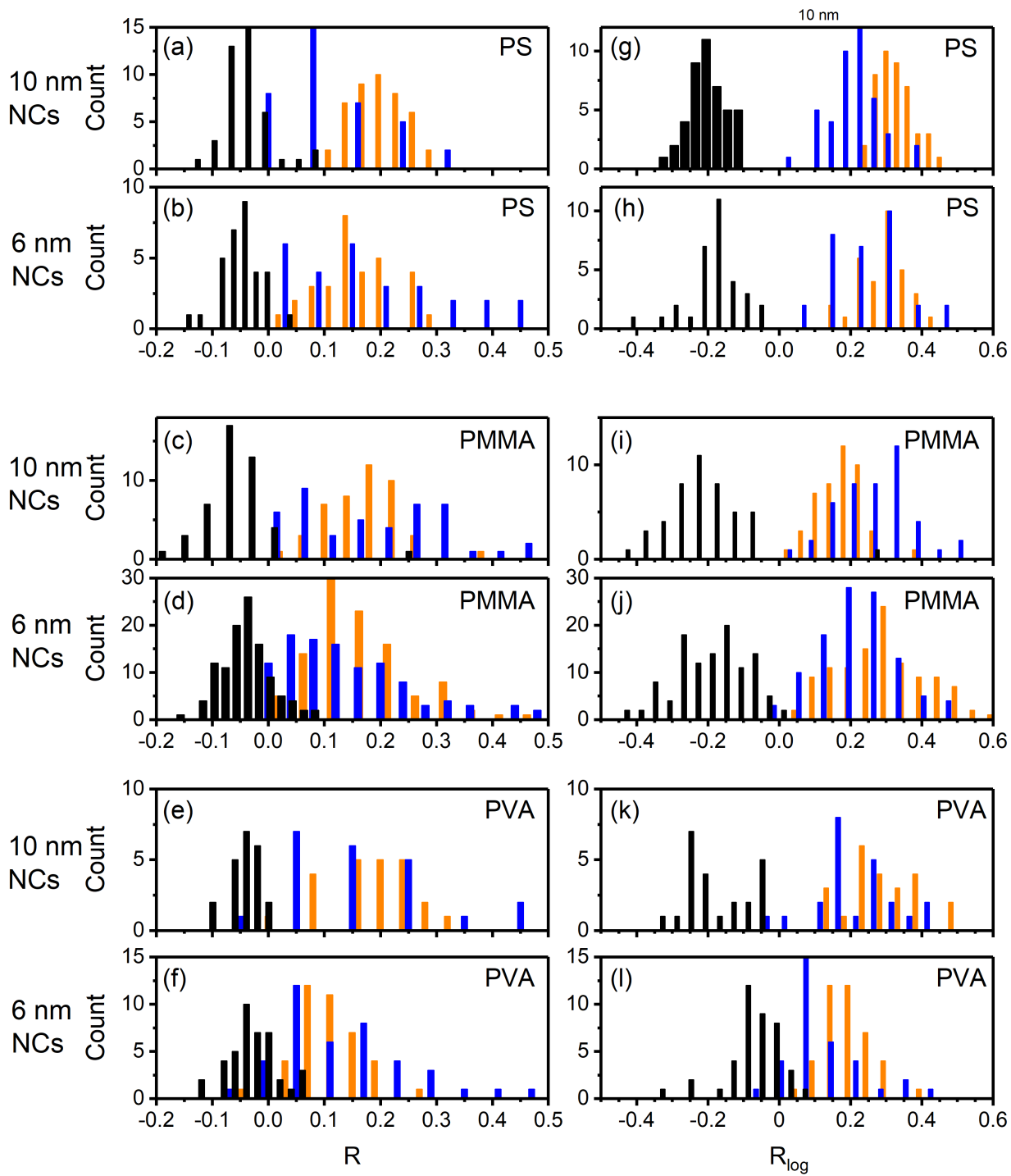

**Figure S13: Correlation coefficient histograms for NCs of various sizes in various polymers.** The average NC sizes are indicated on the left of the figures. (a-f) R values in three polymer matrixes; (g-l) $R_{log}$ values.